\documentstyle[eqsecnum,aps,12pt]{revtex}
\def\s0#1#2{\mbox{\small{$ \frac{#1}{#2} $}}}
\def\0#1#2{\frac{#1}{#2}}
\def\step{{\vspace{.5em}}}

\def\di{\displaystyle}
\def\tab{&\di}
\def\bg{\begin{eqnarray}\begin{array}{rcl}\displaystyle}
\def\eg{\end{array} &\di    &\di   \end{eqnarray}}
\def\bm#1{\begin{eqnarray}\begin{array}{#1}\di}
\def\bmo#1{\begin{eqnarray*}\begin{array}{#1}\di}

\def\bml#1#2{\begin{eqnarray}\begin{array}{#1}\label{#2}\di}

\def\bgo{\begin{eqnarray*}\begin{array}{rcl}\displaystyle}
\def\ego{\end{array} &\di    &\di \nonumber  \end{eqnarray*}}

\def\btensor#1#2{\renew\left#1\begin{array}{#2}\di}
\def\brtensor#1#2#3{\ren#3\left#1\begin{array}{#2}}
\def\botensor#1#2{\renew\left#1\begin{array}{#2}}
\def\etensor#1{\end{array}\right#1}

\def\eq#1{(\ref{#1})}
\def\Eq#1{Eq.~(\ref{#1})}

\def\d{{d}}
\def\tr{{\rm tr}}
\def\Tr{{\rm Tr}}

\def\id{1\!\mbox{l}}

\def\ov{\over}

\def\CO{{\cal O}}
\def\CD{{\cal D}}
\def\CW{{\cal W}}

\def\del{{\mbox{\boldmath$\delta$}}}

\def\ren#1{\renewcommand{\arraystretch}{#1}}

\def\renew{\renewcommand{\arraystretch}{1}}

\makeatletter

\renewenvironment{thebibliography}[1]
         {\section*{References}\frenchspacing\small
          \begin{list}{[\arabic{enumi}]}
         {\usecounter{enumi}\parsep=2pt\topsep 0pt
         \settowidth{\labelwidth}{[#1]}
         \leftmargin=\labelwidth\advance\leftmargin\labelsep
         \rightmargin=0pt\itemsep=0pt\sloppy}}{\end{list}}

\makeatother

\begin{document}
\begin{center}

\thispagestyle{empty}

{\normalsize\begin{flushright}
CERN-TH-2002-049\\
FAU-TP3-02-06\\[10ex] \end{flushright}
}

{\large \bf Renormalisation group flows for gauge theories in axial
  gauges}
\\[6ex]

{Daniel F. Litim
\footnote{Daniel.Litim@cern.ch}
and 
Jan M.~Pawlowski 
\footnote{jmp@theorie3.physik.uni-erlangen.de}}
\\[4ex]
{${}^*${\it 
Theory Division, 
CERN\\
CH-1211 Geneva 23.
}\\[1ex]${}^\dagger${\it 
Institut f\"ur Theoretische Physik III\\
Universit\"at Erlangen, 
D-91054 Erlangen.
}}
\\[10ex]
{\small \bf Abstract}\\[2ex]
\begin{minipage}{14cm}{\small 
    Gauge theories in axial gauges are studied using Exact
    Renormalisation Group flows. We introduce a background field in
    the infrared regulator, but not in the gauge fixing, in contrast
    to the usual background field gauge. We discuss the absence of
    spurious singularities and the finiteness of the flow. It is shown
    how heat-kernel methods can be used to obtain approximate
    solutions to the flow and the corresponding Ward identities.  New
    expansion schemes are discussed, which are not applicable in
    covariant gauges. As an application, we derive the one-loop
    effective action for covariantly constant field strength, and the
    one-loop $\beta$-function for arbitrary regulator.}
\end{minipage}
\end{center}

\newpage
\pagestyle{plain}
\setcounter{page}{1}

\maketitle

\section{Introduction}

The perturbative sector of QCD is very well understood due to the weak
coupling of gluons in the ultraviolet (UV) limit, known as asymptotic
freedom. In the infrared (IR) region, however, the quarks and gluons
are confined to hadronic states and the gauge coupling is expected to
grow large. Thus the IR physics of QCD is only accessible with
non-perturbative methods.  The exact renormalisation group (ERG)
provides such a tool \cite{Wilson,ERG}. It is based on a regularised
version of the path integral for QCD, which is solved by successively
integrating-out momentum modes.\step

ERG flows for gauge theories have been formulated in different ways
(for a review, see \cite{Litim:1998nf}). Within covariant gauges, ERG
flows have been studied in
\cite{Reuter:1994kw,Ellwanger:1994iz,Bonini:1994sj}, while general
axial gauges have been employed in \cite{Litim:1998qi,Litim:1999wk}.
In these approaches, gauge invariance of physical Greens functions is
controlled with the help of modified Ward or Slavnov-Taylor identities
\cite{Ellwanger:1994iz,Bonini:1994sj,Litim:1998qi,Litim:1999wk,D'Attanasio:1996jd,Freire:2000bq}. A different line
has been followed in \cite{Morris:2000px} based on gauge invariant
variables, e.g.~Wilson loops. Applications of these methods to gauge
theories include the physics of superconductors \cite{ScalarQED}, the
computation of instanton-induced effects \cite{Pawlowski:1996ch}, the
heavy quark effective potential
\cite{Ellwanger:1996qf,Bergerhoff:1998cv}, effective gluon
condensation \cite{Reuter:1994yq}, Chern-Simons theory
\cite{Reuter:1996tr}, monopole condensation \cite{Ellwanger:1997wv},
chiral gauge theories \cite{Bonini:1998yv}, supersymmetric Yang-Mills
theories \cite{Falkenberg:1998bg}, and the derivation of the universal
two-loop beta function \cite{Pawlowski:2001df}.\step

In the present paper, we use flow equations to study Yang-Mills
theories within a background field method. In contrast to the usual
background field formalism \cite{abbott}, we use a general axial
gauge, and not the covariant background field gauge. The background
field enters only through the regularisation, and not via the gauge
fixing. Furthermore, in axial gauges no ghost degrees of freedom are
present and Gribov copies are absent.  Perturbation theory in axial
gauges is plagued by spurious singularities of the propagator due to
an incomplete gauge fixing, which have to be regularised separately.
Within an exact renormalisation group approach, and as a direct
consequence of the Wilsonian cutoff, these spurious singularities are
absent \cite{Litim:1998qi}. The resulting flow equation can be used
for applications even beyond the perturbative level.  This formalism
has been used for a study of the propagator \cite{Geiger:1999xj}, for
a formulation of Callan-Symanzik flows in axial gauges
\cite{Simionato:2000iz}, and for a study of Wilson loops
\cite{Panza:2000tg,Panza:2001dg}.\step

Here, we continue the analysis of \cite{Litim:1998qi,Litim:1999wk} and
provide tools for the study of Yang-Mills theories within axial
gauges. First we detail the discussion of the absence of spurious
singularities. Then a framework for the evaluation of the path
integral for covariantly constant fields is discussed. We use an auxiliary
background field which allows us to define a gauge invariant effective
action. The background field is introduced only in the regulator, in
contrast to the usual background field formalism.  This way it is
guaranteed that all background field dependence vanishes in the
infrared limit, where the cutoff is removed. We employ heat kernel
techniques for the evaluation of the ERG flow. The heat kernel is used
solely as a technical device, and not as a regularisation. The flow
equation itself is by construction infra-red and ultra-violet finite
and no further regularisation is required.  As an explicit
application, we compute the full one-loop effective action for
non-Abelian gauge theories. This includes the universal
$\beta$-function at one loop for arbitrary regulator. We also discuss
new expansions of the flow, which are not applicable for covariant
gauges.  \step

The work is organised as follows. We begin with a brief review of the
Wilsonian approach for gauge theories. This includes a derivation of
the flow equation. We discuss the absence of spurious singularities
and the finiteness of the flow. This leads to a mild restriction on
the fall-off behaviour of regulators at large momenta.  (Section
\ref{ERG}). Next, we consider the implications of gauge symmetry. This
includes a discussion of the Ward-Takahashi identities, the
construction of a gauge-invariant effective action, and the study of
the background field dependence. Explicit examples for background
field dependent regulators are also given (Section
\ref{sec:Symmetries}). We derive the propagator for covariantly
constant fields, and explain how expansions in the fields and heat
kernel techniques can be applied in the present framework (Section
\ref{Methods}). We compute the full one loop effective action using
heat kernel techniques. We also show in some detail how the universal
beta function follows for arbitrary regulator functions (Section
\ref{Applications}). We close with a discussion of the main results
(Section \ref{Discussion}) and leave some more technical details to
the Appendices.

\section{Wilsonian approach  for gauge theories} \label{ERG}

In this section we review the basic ingredients and assumptions
necessary for the construction of an exact renormalisation group
equation for non-Abelian gauge theories in general axial gauges. 
This part is based on earlier work \cite{Litim:1998qi,Litim:1999wk}.
New material is contained in the remaining subsections, where we
discuss the absence of spurious singularities and the finiteness of
the flow.

\subsection{Derivation of the flow} 

The starting point for the derivation of an exact renormalisation
group equation are the classical action $S_A$ for a Yang-Mills theory,
an appropriate gauge fixing term $S_{\rm gf}$ and a regulator term
$\Delta S_k$, which introduces an infra-red cut-off scale $k$
(momentum cut-off). This leads to a $k$-dependent effective action
$\Gamma_k$. Its infinitesimal variation w.r.t.\ $k$ is described by
the flow equation, which interpolates between the gauge-fixed
classical action and the quantum effective action, if $\Delta S_k$ and
$\Gamma_k$ satisfy certain boundary conditions at the initial scale
$\Lambda$.  The classical action of a non-Abelian gauge theory is
given by
\begin{eqnarray}\label{classical}
S_A[A]=\frac{1}{4}\int d^4 x\,
 F_{\mu\nu}^a(A) F_{\mu\nu}^a (A)
\end{eqnarray}
with the field strength tensor 
\begin{eqnarray}\label{Fmunu}
F_{\mu\nu}^a(A)=\partial_\mu A^a_\nu-\partial_\nu A^a_\mu + g f^a_{\ \ 
  bc} A^b_\mu A^c_\nu 
\end{eqnarray} 
and the covariant derivative 
\begin{eqnarray}\label{covder}
D^{ab}_\mu(A)=\delta^{ab}\partial_\mu + g f^{acb}A^c_\mu, \qquad  
[t^b,t^c]={f_a}^{bc}t^a.  
\end{eqnarray}
A general axial gauge fixing is given by
\begin{eqnarray}\label{axial}
 S_{\rm gf}[A]=
{1\ov 2}\int d^4 x\, n_\mu A^a_\mu\ {1\ov \xi n^2}\ n_\nu A^a_\nu.  
\end{eqnarray} 
The gauge fixing parameter $\xi$ has the mass dimension $-2$ and may
as well be operator-valued \cite{Litim:1998qi}. The particular
examples $\xi=0$ and $\xi p^2=-1$ are known as the axial and the
planar gauge, respectively. The axial gauge is a fixed point of the
flow \cite{Litim:1998qi}.\step
          
The scale-dependent regulator term is
\begin{eqnarray}\label{deltaS}
\Delta S_k [A,\bar A]= \frac{1}{2} \int d^4x A^a_\mu
\left. R_k\right.^{ab}_{\mu\nu}[\bar A] A^b_\nu\ .  
\end{eqnarray}
It is quadratic in the gauge field and leads to a modification of the
propagator.  We have introduced a background field $\bar A$ in the
regulator function. Both the classical action and the gauge fixing
depend only on $A$.  The background field serves as an auxiliary field
which can be interpreted as an index for a family of different
regulators $R_{k,\bar A}$. Its use will become clear below.  \step

The scale dependent Schwinger functional $W_k[J,\bar A]$, given by
\begin{eqnarray}\label{Schwingerk} \exp W_k[J,\bar A] =
\int {\CD}A \exp\left\{-S_k[A,\bar A]
+ \int d^4 x\, A^a_\mu J^a_\mu\right\}, 
\end{eqnarray} 
where 
\begin{eqnarray}\label{Sk} 
S_k[A,\bar A]=S_A[A]+S_{\rm
  gf}[A]+\Delta S_k [A,\bar A]\,. 
\end{eqnarray}
We introduce the scale dependent effective action $\Gamma_k[A,\bar A]$
as the Legendre transform of (\ref{Schwingerk})
\begin{eqnarray}\label{Gkdef} \Gamma_k[A,\bar A] = 
\int d^4 x J^a_\mu A^a_\mu -W_k[J,\bar A]-\Delta
S_k[A,\bar A],\quad A^a_\mu 
= \frac{\delta W_k[J,\bar A]}{\delta J^a_\mu}.
\end{eqnarray} 
For later convenience, we have subtracted $\Delta S_k $ from the
Legendre transform of $W_k$. Thus $\Gamma_k[A,\bar A]$ is given by the
integro-differential equation
\begin{eqnarray}\label{barG}
\exp -\Gamma_k[A,\bar A]= 
\int \CD a\, \exp\Bigl
\{-S_A[a]-S_{\rm gf}[a]-\Delta S_k [a-A,\bar A]+ {\delta 
\ov \delta A}\Gamma_k[A,\bar A](a-A)\Bigr\}. 
\end{eqnarray} 
The corresponding flow equation for the effective action
\begin{eqnarray}\label{flow} 
\partial_t\Gamma_k[A,\bar A]=\frac{1}{2}\Tr 
\left\{G_k[A,\bar A]\, \partial_t R_k[\bar A] \right\} 
\end{eqnarray}
follows from \eq{barG} by using $\langle a-A\rangle=0$.  The trace
sums over all momenta and indices, $t=\ln k$. $G_k$ is the full
propagator of the field $A$, whereas $\bar A$ is not propagating.
Its inverse is given by
\begin{eqnarray}\label{G_k} 
\left. \left(G_k[A,\bar A]\right)^{-1}\,\right.^{ab}_{\mu\nu}(x,x')= 
\frac{\delta^2\Gamma_k[A,\bar A]}{\delta
  A^\mu_a(x)\delta A^\nu_b(x')}+\left.
R_k[\bar A]\,\right.^{ab}_{\mu\nu}(x,x'). 
\end{eqnarray}   
There are no ghost terms present in \eq{flow} due to the axial gauge
fixing. For the regulator $R_k$ we require the following properties at 
$\bar A=0$. 
\begin{eqnarray}\label{Regulator}
\lim_{p^2/k^2 \to \infty} p^2 R_k =0, 
\qquad \qquad \lim_{p^2/k^2\to 0} R_k\sim p^2 
\left(\frac{k^2}{p^2}\right)^\gamma, 
\end{eqnarray} 
where $p^2$ is plain momentum squared. 
Regulators with $\gamma=1$ have a mass-like infra-red behaviour with
$R_k(0)\sim k^2$. The example in \eq{furnish} has $\gamma=1$.  In
turn, regulator with $\gamma > 1$ diverge for small momenta. The latter
condition in \eq{Regulator} implies that $R_k$ introduces an IR
regularisation into the theory. The first condition in \eq{Regulator}
ensures the UV finiteness of the flow in case that $G_k\propto p^{-2}$
for large $p^2$. For covariant gauges this is guaranteed. Within axial
gauges, additional care is necessary because of the presence of
spurious singularities. 
It is seen by inspection of \eq{barG} and \eq{Regulator} that the
saddle-point approximation about $A$ becomes exact for $k\to \infty$.
Here, $\Gamma_k$ approaches the classical action.  For $k\to 0$, in
turn, the cut-off term disappears and we end up with the full quantum
action. Hence, we confirmed that the functional $\Gamma_k$ indeed
interpolates between the gauge-fixed classical and the full quantum
effective action:
\begin{mathletters}\label{interpolate}
\begin{eqnarray}
\label{UV-Limit}
\lim_{k\to\infty}\Gamma_k[A,\bar A]&\equiv &S[A]+S_{\rm gf}[A], 
\\
\label{IR-Limit}
\lim_{k\to 0}\Gamma_k[A,\bar A]&\equiv &\Gamma[A]. 
\end{eqnarray} 
\end{mathletters}%
Notice that both limits are independent of $\bar A$ supporting the 
interpretation of $\bar A$ as an index for a class of flows.
It is worth emphasising that both the infrared and ultraviolet 
finiteness of \eq{flow} are ensured by the conditions \eq{Regulator} 
on $R_k$.  

\subsection{Absence of spurious singularities}\label{sec:absence} 
The flow equation \eq{flow} with a choice for the initial effective
action $\Gamma_\Lambda$ at the initial scale $\Lambda$ serves upon
integration as a definition of the full effective action
$\Gamma=\Gamma_{k=0}$.  It remains to be shown that \eq{flow} is
finite for all $k$ thus leading to a finite $\Gamma$. In particular
this concerns the spurious singularities present in perturbation
theory: the propagator $P_{\mu\nu}$ related to $S=S_A+S_{\rm gf}$ is
\begin{eqnarray}\label{eq:Prop1}
P_{\mu\nu}=\frac{\delta_{\mu\nu}}{p^2}+\frac{n^2(1+\xi p^2)}{
  (n p)^2}\frac{p_\mu p_\nu}{p^2}
-\frac{1}{p^2}\frac{\left(n_\mu p_\nu +n_\nu p_\mu\right)}{ n
  p}.  
\end{eqnarray} 
It displays the usual IR poles proportional to $1/p^2$. We observe
additional divergences for momenta orthogonal to $n_\mu$.  These poles
appear explicitly up to second order in $1/(np)$ and can even be of
higher order for certain $(np)$-dependent choices of $\xi$.  For the
planar gauge $\xi p^2=-1$, the spurious divergences appear only up to
first order.\step

This artifact makes the application of perturbative techniques very
cumbersome as an additional regularisation for these spurious
singularities has to be introduced. We argued in \cite{Litim:1998qi}
that these spurious singularities are missing in the flow equation.
Here, we further the discussion, also providing some information about
the intricate limit where the cut-off is removed. First of all we
derive a bound on the flow \eq{flow}. Then, we argue that this bound
results in weak constraints on the decay behaviour of the regulator
function $r$ for large momenta. This is sufficient for providing a
well-defined RG flow.\step

We start with an analysis of the momentum dependence of the propagator
in the presence of the regulator. To that end we set the background
field to zero, $\bar A=0$, and specify the regulator as
\begin{eqnarray}
{R}^{ab}_{k,\mu\nu}(p)=\delta^{ab}\left[r(p^2)p^2\delta_{\mu\nu}-\tilde
  r(p^2) p_\mu p_\nu\right].  
\label{eq:cut-off2}
\end{eqnarray}
The IR/UV limits of $r,\tilde r$ can be read-off from \eq{Regulator}.
In \eq{eq:cut-off2} we did not introduce terms with tensor structure $
(n_\mu p_\nu+n_\nu p_\mu)$ and $n_\mu n_\nu$.  For the present
purpose, the discussion of spurious singularities, the choice
\eq{eq:cut-off2} suffices. Indeed, even $\tilde r$ plays no r$\hat{\rm
  o}$le for the absence of spurious singularities in the flow equation
approach. The only important term for the discussion of spurious
singularities is that proportional to the term 
$p^2\delta_{\mu\nu}\delta^{ab}$. 
It is this term, proportional to the identity operator, that
guarantees the suppression of all momentum modes for large cut-off.
The other tensor structures are proportional to projection operators
and cannot lead to a suppression of all modes. With a regulator
obeying \eq{eq:cut-off2} the propagator takes the form
\begin{eqnarray}
P_{k,\mu\nu}=a_1\frac{\delta_{\mu\nu}}{p^2}+a_2 \frac{p_\mu 
p_\nu}{p^4} +a_3
\frac{n_\mu p_\nu +n_\nu p_\mu}{p^2 (np)}+a_4 \frac{n_\mu
  n_\nu}{n^2 p^2},
\label{eq:prop_k}
\end{eqnarray} 
with the dimensionless coefficients 
\begin{mathletters}
\begin{eqnarray} \di a_1 &=&
  1/(1+r)\,,\\
a_2 &=& (1+\tilde r)(1 + 
\xi p^2 (1+r))/z \,,\\
a_3 &=& -(1+\tilde r) s^2/z\,, \\
 a_4 &=& -(r-\tilde r)/z, 
\end{eqnarray} 
and
\begin{eqnarray} 
s^2& =& (np)^2/(n^2 p^2),\\ z& =& (1+r)[(1+\tilde r) s^2 +(r-\tilde
r)(1+p^2\xi (1+r))].  \end{eqnarray} 
\end{mathletters}%
Now we evaluate the different limits in $p^2$ and $k$ important for
the approach. To keep things simple we restrict ourselves to the case
$\tilde r=0$ and a regulator $r$ leading to a mass-like IR behaviour:
$\lim_{ p^2/k^2 \to 0} r(p^2)=k^2/p^2$.  For this choice we deduce
from \eq{eq:prop_k} and \eq{Regulator} that $P_{k,\mu\nu}$ has the
limits
\begin{eqnarray} 
\lim_{p^2/k^2\rightarrow \infty}P_{k,\mu\nu} = P_{\mu\nu}, & \qquad \qquad 
& 
\lim_{p^2/k^2\rightarrow 0}P_{k,\mu\nu} = \frac{1}{k^2}\left(
  \delta_{\mu\nu}+\frac{n_\mu n_\nu}{n^2}
  \frac{1}{1+\xi k^2}\right), 
\end{eqnarray}  
with $P_{\mu\nu}$ defined in \eq{eq:Prop1}. 
By construction, the propagator \eq{eq:prop_k} is IR
finite for any $k>0$. Now, the important observation is the following:
in contrast to the perturbative propagator $P_{\mu\nu}$, the limit of
$P_{k,\mu\nu}$ for $np\to 0$ is finite. This holds true even
 for an arbitrary choice of $\xi(p,n)$ and leads to 
\begin{eqnarray}
P_{k,\mu\nu}  = \di \frac{1}{1+r}\frac{\delta_{\mu\nu}}{p^2}+
\frac{1}{(1+r)r}\frac{p_\mu p_\nu}{p^4}
-\di\frac{1}{(1+r)(1+p^2\xi (1+r))} \frac{n_\mu n_\nu}{n^2
  p^2}\label{eq:prop_np}.  
\end{eqnarray}
Thus (\ref{eq:prop_np}) is well-behaved and finite for all momenta
$p$. The plain spurious divergences are already absent as soon as the
infra-red behaviour of the propagator is under control. This 
holds for $R$ with the most general tensor structure as long as it obeys 
the limits \eq{Regulator}. For example, it is easily
extended to non-zero $\tilde r$ as long as the regulators $r$ and
$\tilde r$ have not been chosen to be identical. Already in the
infrared region $\tilde r$ has to be smaller than $r$ in order to have
a suppression of longitudinal modes at all. So we discard the option
of identical $r$ and $\tilde r$. \step 

Still, for $n p=0$ and large momenta squared $y=p^2$ the regulator
tends to zero and the second term in \eq{eq:prop_np} diverges in the
limit $y\to \infty$ proportional to $y^{-1} (r-\tilde r)^{-1}>
y^{d/2-1}$, following from \eq{Regulator}. Hence, even though the term
only diverges for $y\to \infty$, a more careful analysis is needed for
proving the finiteness of the flow equation. We emphasise that the
remaining problem is the integration over {\it large momenta} in the
flow equation and {\it not} an IR problem at vanishing momentum. Thus,
by showing that this problem is absent in the flow equation for all
$k$ it cannot reappear at $k=0$. Indeed, we shall see that finiteness
of the flow for all $k$ implies a stronger decay of the regulator for
large momenta as in \eq{Regulator}.  In turn, one may expect problems
for regulators with weaker decay. \step

\subsection{Finiteness of the RG flow}\label{sec:finite} 

Here, finiteness of the flow equation is proven by deriving an upper
bound for the flow following a bootstrap approach. The derivation of
the flow equation is based on the existence of a finite renormalised
Schwinger functional for the full theory \cite{Pawlowski:2001df}. In
the present context this only implies the {\it existence} of a
renormalisation procedure for axial gauges, the form of which is then
determined by integrating the flow. An explicit systematic
constructive renormalisation procedure is not required. The latter is
a problem in perturbative field theory: no renormalisation procedure
is known, which can be proven to be valid to all orders of
perturbation theory.  \step

In the present approach, spurious singularities could spoil the
finiteness of \eq{flow} due to infinities arising from the integration
of the large momentum domain. For the derivation of a bound we can
safely assume, that for all $k$ and large momenta $p^2$ the full
propagator $\Gamma_k^{(2)}$ is dominated by its classical part
(possibly with some multiplicative renormalisation constants). Hence
for large momenta we can estimate $\Gamma^{(2)}_k (S^{(2)}+S_{\rm
  gf}^{(2]})^{-1} < C[A,\bar A]$ with $C[A,\bar A]>0$.  Consequently
the field independent part of the flow provides a bound on the full
flow. The only terms that could produce divergences are related to the
terms in \eq{eq:prop_k} proportional to $a_2$ and $a_3$, the source
for divergences being $z^{-1}$. The coefficient $a_4$ of the last term
in \eq{eq:prop_k} also contains $z^{-1}$ but also an additional factor
$r$. Hence the limit $np\to 0$ can be safely done in the term
$a_4$.\step

We do not go into the details of the computation. A more detailed
derivation and discussion is given elsewhere. We quote the result for
$\tilde r=0$. Upon integrating the angular $s$-part of the momentum
integration we get an estimate from the part of $ \Tr\,
P_k\,\partial_t R_k$ with the slowest decay for $y\to\infty$
\begin{eqnarray}\label{eq:estimate} 
{\rm bound} \propto \left|\int_{a}^\infty 
d y\, y^{2}\0{\sqrt{1+y\xi}}{1+a\xi}\, 
\0{r'(y)}{\sqrt{r(y)}}\right|\,, 
\end{eqnarray} 
where the square root terms stem from an integration $\int_{-1}^1
ds/[s^2+(1+\xi y\, )r(y)]$. Since the potential problem only occurs
from an integration over large momenta squared $y=p^2$, we have
restricted the $y$-integral to $y\geq a$ where $a$ is at our disposal.
It can be chosen the same for all $k$. This ensures that the limit
$k\to 0$ can be taken smoothly. The bound \eq{eq:estimate} stems from
the second term in \eq{eq:prop_k} proportional to $a_2$.
\Eq{eq:estimate} is finite for regulators $r$ that decay faster than
$y^{-5}$. Without spurious singularities, $r$ has to decay stronger
than $y^{-2}$, see \eq{Regulator}. Hence we have a mild additional
constraint due to the fact that the full propagator $G_k$ does not
introduce an additional suppression. Typically, the regulator is
chosen to decay exponentially for large momenta. Similar finite
integrals as in \eq{eq:estimate} also occur in field dependent terms
in the flow, as we shall see later in Sect.~\ref{Applications}. \step

This analysis shows the finiteness of the flow \eq{flow} and supports
the claim that the flow equation provides a consistent quantisation
procedure for gauge theories in axial gauges.  The bound also marks
the use of Callan-Symanzik (CS) type flows ($R_k\propto k^2$ and
$r(y)\propto y^{-1}$) as questionable in axial gauges. It has been
already mentioned in \cite{Litim:1998nf} that such a choice requires
an additional renormalisation. The presence of contributions from all
momenta at every flow step makes the limit $k\to 0$ an extremely
subtle one. This limit is very sensitive to a proper fine-tuning. In
axial gauges, this problem for CS flows gets even worse due to the
spurious singularities. We know that a consistent renormalisation
procedure in the axial gauge is certainly non-trivial.  For CS-type
flows, one is back to the original problem of spurious singularities
in perturbation theory, but with a more difficult propagator and
additional renormalisation problems. A recent calculation of
perturbative corrections to the Wilson loop has indeed shown that
formulations in axial gauge with a mass term for the gauge field meet
problems \cite{Panza:2000tg,Panza:2001dg}. The massless limit of this
observable did not coincide with the well-known result.  In turn, for
regulators which decay faster than $r(y)\sim y^{-5}$, the problem is
cured.\step

\section{Symmetries}\label{sec:Symmetries}

In this section, the issue of gauge invariance of physical Greens
functions, controlled by modified Ward-Takahashi identities, is
studied.  We discuss the role of background fields, which, in
contrast to the usual background field method
\cite{abbott,Litim:1999wk}, will only be introduced for the Wilsonian
regulator term. The Ward-Takahashi identities for the quantum and the
background field are derived. We define a gauge-invariant effective
action as it follows from the present formalism, and discuss its
background field dependence. Finally, we discuss the background field
dependent regularisation.

\subsection{Modified Ward-Takahashi Identities}\label{secmWI}

We now address the issue of gauge invariance for physical Greens
functions. The problem to face is that the presence of a regulator
term quadratic in the gauge fields is, a priori, in conflict
with the requirements of a (non-linear) gauge symmetry.  This question
has been addressed earlier for Wilsonian flows within covariant gauges
\cite{Reuter:1994kw,Ellwanger:1994iz,Bonini:1994sj,Litim:1998qi,D'Attanasio:1996jd}.
The resolution to the problem is that modified Ward-Takahashi
identities (as opposed to the usual ones) control the flow such that
physical Greens functions, obtained from $\Gamma_k$ at $k=0$,
satisfy the usual Ward-Takahashi identities.\step

The same line of reasoning applies in the present case even though in
the presence of the background field $\bar A$ some refinement is
required \cite{Litim:1999wk}.  In this particular point it is quite
similar to the symmetry properties of the full background field
formalism as discussed in \cite{Freire:2000bq}. The background field
makes it necessary to deal with two kinds of modified Ward-Takahashi
Identities. The first one is related to the requirement of gauge
invariance for physical Green functions, and is known as
 modified Ward Identity (mWI).  The second one has to do with the
presence of a background field $\bar A$ in the regulator term $R_k$,
and will be denoted as the background field Ward-Takahashi
Identity (bWI).\step

To simplify the following expressions let us introduce the
abbreviation $\del_\omega$ and $\bar\del_\omega$ for the generator of
gauge transformations on the fields $A$ and $\bar A$ respectively:
\begin{mathletters}\label{trafo} 
\begin{eqnarray}\label{trafo1} 
\del_\omega A\tab =\tab D(A)\omega \qquad  \del_\omega \bar A=0
\\
\di  \bar\del_\omega A\tab =\tab 0 \qquad \qquad  
\ \ \bar\del_\omega \bar A = D(\bar A)\omega. 
\label{trafo2}\end{eqnarray}
\end{mathletters}%
The action of the gauge transformations $\del_\omega$ and
$\bar\del_\omega$ on the effective action $\Gamma_k$ can be computed
straightforwardly.  It is convenient to define
\begin{mathletters}\label{Wdef}
\begin{eqnarray}\label{W1}
\CW_k[A,\bar A;\omega] &\equiv& \di 
\del_\omega\Gamma_k[A,\bar A] 
-\Tr\,(n_\mu \partial_\mu \omega)\ \frac{1}{n^2\xi} n_\nu 
A_\nu 
+\frac{1}{2}\Tr\,\omega \left[G_{k}[A,\bar A],
R_k[\bar A]\right]\\\di 
\bar\CW_k[A,\bar A;\omega]&\equiv& \bar\del_\omega 
\Gamma_{k}[A,\bar A] -\frac{1}{2}\Tr\,\omega\left[ 
G_{k}[A,\bar A],R_k[\bar A] \right]\ . 
\label{W2}
\end{eqnarray}
\end{mathletters}
In terms of \eq{Wdef}, the behaviour of $\Gamma_k[A,\bar A]$ under the
transformations $\del_\omega$ and $\bar\del_\omega$, respectively, is
given by
\begin{mathletters}\label{WIs}   
\begin{eqnarray}\label{mWI}
\CW_k[A,\bar A;\omega]\tab =\tab 0 \\\di 
\bar\CW_k[A,\bar A;\omega]\tab =\tab 0 \label{bWI}
\end{eqnarray}
\end{mathletters}%
\Eq{bWI} is valid for regulators $R_k$  that transform as tensors 
under $\del_\omega$,
\begin{eqnarray}\label{trafofR}
\bar\del_\omega R_k[\bar A]=\left[R_k[\bar A],\omega\right]\ .
\end{eqnarray} 
\Eq{mWI} is referred to as the modified Ward-Takahashi identity, and
\eq{bWI} as the background field Ward-Takahashi identity.\step

Let us show that \eq{WIs} is consistent with the basic flow equation
\eq{flow}. With consistency, we mean the following. Assume, that a
functional $\Gamma_k$ is given at some scale $k$ which is a solution
to both the mWI and the bWI. We then perform a small integration step
from $k$ to $k'=k-\Delta k$, using the flow equation, and ask whether
the functional $\Gamma_{k'}$ again fulfils the required Ward
identities \eq{WIs}.  That this is indeed the case is encoded in the
following flow equations for \eq{WIs}, namely
\begin{mathletters}\label{compatible}  
\begin{eqnarray}\label{compatible1}
\partial_t \CW_k[A,\bar A;\omega]\tab = \tab 
-\frac{1}{2}{\rm Tr}\left( G_k \frac{\partial
    R_k}{\partial t} G_k \frac{\delta}{\delta
    A}\otimes\frac{\delta}{\delta A}\right)\CW_k[A,\bar A;\omega]\\\di   
\partial_t \bar\CW_k[A,\bar A;\omega] \tab =\tab  
\frac{1}{2}{\rm Tr}\left( G_k \frac{\partial
    R_k}{\partial t} G_k \frac{\delta}{\delta
    A}\otimes\frac{\delta}{\delta A}\right)\bar\CW_k[A,\bar A;\omega], 
\label{compatible2}
\end{eqnarray}
\end{mathletters}%
where $\left(\frac{\delta}{\delta A}\otimes\frac{\delta}{\delta A}
\right)_{\mu\nu}^{ab}(x,y)= \frac{\delta}{\delta
  A^\mu_a(x)}\frac{\delta}{\delta A^\nu_b(y)}$.  \Eq{compatible}
states that the flow of mWI is zero if the mWI is satisfied for the
initial scale. The required consistency follows from the fact that the
flow is proportional to the mWI itself \eq{compatible1}, which
guarantees that \eq{mWI} is a fixed point of \eq{compatible1}.  The
same follows for the bWI by using \eq{compatible2}.  There is no
fine-tuning involved in lifting a solution to \eq{mWI} to a solution
to \eq{bWI}. It also straightforwardly follows from \eq{compatible1}
and \eq{compatible2}.\step

We close with a brief comment on the use of mass term regulators. Such
a regulator corresponds simply to $R_k=k^2$ and leads to a
Callan-Symanzik flow. The regulator is momentum-independent, which
implies that the loop term in \eq{W1} vanishes identically.  Hence one
concludes that the modified Ward identity reduces to the usual one for
all scales $k$. This happens only for an axial gauge fixing
\cite{Litim:1998qi}.

\subsection{Gauge invariant effective action}

Returning to our main line of reasoning and taking advantage of the
results obtained in the previous section, we define a gauge invariant
effective action only dependent on $A$ by identifying $\bar A=A$.  It
is obtained for a particular choice of the background field, and
provides the starting point for our formalism.\step

It is a straightforward consequence of the mWI \eq{mWI} and the bWI
\eq{bWI} that the effective action $\Gamma_k[A,\bar A]$ is gauge
invariant -- up to the gauge fixing term -- under the {combined}
transformation
\begin{eqnarray}\label{gaugeinv}
(\del_\omega+\bar\del_\omega)\Gamma_k[A,\bar A]= 
\Tr\, n_\mu (\partial_\mu\omega) 
{1\ov n^2 \xi} n_\nu A_\nu. 
\end{eqnarray} 
We define the effective action $\hat\Gamma_k[A]$ as
\begin{eqnarray}\label{Gk} 
\hat\Gamma_k[A]=\Gamma_k[A,\bar A=A]. 
\end{eqnarray} 
The action $\hat\Gamma_k[A]$ is gauge invariant up to
 the gauge fixing term, to wit
\begin{eqnarray}\label{Gkinv}
\del_\omega \hat\Gamma_k[A]= \Tr\, \left\{n_\mu (\partial_\mu\omega) 
{1\ov n^2 \xi} n_\nu A_\nu\right\}.
\end{eqnarray}
This follows from \eq{gaugeinv}.  Because of \eq{IR-Limit}, the
effective action $\hat\Gamma_{k=0}[A]$ is the full effective action.
The flow equation for $\hat \Gamma_k[A]$ can be read off from the
basic flow equation \eq{flow},
\begin{eqnarray}\label{flowphys} 
\partial_t\hat\Gamma_k[A]=\frac{1}{2}\Tr 
\left\{G_k[A,A]\, \partial_t R_k[A] \right\}, 
\end{eqnarray}
Notice that the right-hand side of \eq{flowphys} is {not} a functional
of $\hat\Gamma_k[A]$. The flow depends on the full propagator
$G_k[A,A]$, which is the propagator of $A$ in the background of $\bar
A$ taken at $\bar A=A$.  Thus for the flow of $\hat\Gamma_k[A]$ one
needs to know the flow (of a subset) of vertices of
$\delta^2\Gamma_k[A,\bar A]/(\delta A)^2$ at $\bar A=A$. Still,
approximations, where this difference is neglected are of some
interest \cite{Litim:2002hj}.  \step

We argue that \eq{Gkinv} has far reaching consequences for the
renormalisation procedure of $\hat\Gamma_k[A]$ as is well-known for
axial gauges and the background field formalism.  $\Gamma_k[A]$ is
gauge invariant up to the breaking due to the gauge fixing term. We
define its gauge invariant part as
\begin{mathletters}\label{gaugeinvan}
\begin{eqnarray}
&&\Gamma_{k,\rm inv}[A]= \Gamma_k[A]-S_{\rm gf}[A]\\
&&\del_\omega\Gamma_{k,\rm inv}[A]=0\ .  
\end{eqnarray}
\end{mathletters}%
\Eq{gaugeinvan} implies that the combination $gA$ is invariant under
renormalisation, $\partial_t (gA)=0$. If one considers wave function
renormalisation and coupling constant renormalisation for $A$ and $g$
respectively
\begin{mathletters}\label{Rfactors}  
\begin{eqnarray}\label{wave} 
A\tab \to \tab Z_F^{1/2} A\\\di 
g\tab \to \tab Z_g g\label{Rg} 
\end{eqnarray}
\end{mathletters} 
we conclude that 
\begin{eqnarray}\label{nonrenorm} 
Z_g = Z_F^{-1/2}\,. 
\end{eqnarray}

\subsection{Background field dependence}\label{bfd}

By construction, the effective action $\Gamma_k[A,\bar A]$ at some
finite scale $k\neq 0$ will depend on the background field $\bar A$.
This dependence disappears for $k=0$.  The effective action $\hat
\Gamma_k[A]$ is the simpler object to deal with as it is gauge
invariant and only depends on one field. As we have already mentioned
below \eq{flowphys}, its flow depends on the the propagator
$\delta^2_A \Gamma_k[A,\bar A]$ at $A=\bar A$. Eventually we are
interested in approximations where we substitute this propagator by
$\delta^2_A\hat\Gamma_k$. The validity of such an approximation has to
be controlled by an equation for the background field dependence of
$\Gamma_k[A,\bar A]$.  The flow of the background field dependence of
$\Gamma_k[A,\bar A]$ can be derived in two ways. $\delta_{\bar A}
\partial_t\Gamma_k$ can be derived from the flow equation \eq{flow},
\begin{eqnarray}\label{bar A1}
\frac{\delta }{\delta \bar A}\partial_t\Gamma_k[A,\bar A]=
\frac{1}{2}{\delta\ov \delta 
\bar A}\, \Tr\,\left\{G_k[A,\bar A]\partial_t 
R_k[\bar A]\right\}\ . 
\end{eqnarray} 
The flow $\partial_t\delta_{\bar A}\Gamma_k$ follows the observation
that the only background field dependence of $\Gamma_k$ originates in
the regulator. Thus, $\delta_{\bar A}\Gamma_k$ is derived along the
same lines as the flow itself and we get
\begin{eqnarray}\label{bar A}
\partial_t\frac{\delta }{\delta \bar A}\Gamma_k[A,\bar A]=
\frac{1}{2}\Tr\,\partial_t \left\{G_k[A,\bar A]{\delta\ov \delta 
\bar A} R_k[\bar A] \right\}, 
\end{eqnarray} 
which turns out to be important also for the derivation of the
universal one loop $\beta$-function in Sect.~\ref{1loop}.  
The difference of \eq{bar A1} and \eq{bar A} has to vanish 
\begin{equation}\label{consistency}
\bigl[{\delta\ov \delta \bar A}\,,\,\partial_t\bigr]\, \Gamma_k[A,\bar A]=0\ .
\end{equation}
\Eq{consistency} combines the flow of the intrinsic 
$\bar A$-dependence of $\Gamma_k[A,\bar A]$ \eq{bar A} with the 
$\bar A$-dependence of the flow equation itself \eq{bar A1}. 
It provides a check 
for the validity of a given approximation. Using the right hand sides of 
\eq{bar A1} and \eq{bar A} the consistency condition \eq{consistency}
can be turned into
\begin{eqnarray}\label{diffbarA}
\Tr\left\{ 
G_k\,{\delta\Gamma_k^{(2)}\ov \delta\bar A}
G_k\, \partial_t R_k\right\}
=\Tr\left\{ G_k\, 
{\delta R_k\ov \delta\bar A}\, G_k 
\partial_t\Gamma_k^{(2)}\right\}, 
\end{eqnarray} 
where 
\begin{eqnarray}\label{2derGk}
\left.\Gamma_k^{(2)}[A,\bar A]\right._{\mu\nu}^{ab}(x,x')=
{\delta^2 \Gamma_k[A,\bar A]\ov 
\delta A_a^\mu(x)\delta A_b^\nu(x')}\,. 
\end{eqnarray}
With \eq{diffbarA}, we control the approximation 
\begin{eqnarray}\label{approx} 
\left.{\delta^2\Gamma_k[A,\bar A]\ov \delta A \ \delta A}\right|_{\bar A=A}= 
{\delta^2\hat\Gamma_k[A]\ov \delta A \ \delta A}+\mbox{\rm sub-leading\ terms} 
\end{eqnarray} 
For this approximation the flow \eq{flowphys} is closed and can be
calculated without the knowledge of $\Gamma_k^{(2)}$, but with
$\hat\Gamma_k^{(2)}$. Amongst others, the approximation \eq{approx} is
implicitly made within proper-time flows, where the use of heat-kernel
methods is even more natural \cite{Liao:1995nm}.  This is discussed in
\cite{Litim:2001hk} (see also \cite{Litim:2002hj}).  Let us finally
comment on the domain of validity for the approximation \eq{approx}.
In the infrared $k\to 0$, the dependence of the effective action
$\Gamma_k[A,\bar A]$ on the background field $\bar A$ becomes
irrelevant, because the regulator $R_k[\bar A]$ tends to zero.
Therefore we can expect that \eq{approx} is reliable in the infrared,
which is the region of interest.

\subsection{Regulators}

We have seen that the symmetries of the effective action $\Gamma_k$
and the flow crucially depend on the properties of $R_k[\bar A]$, in particular the construction of a gauge invariant effective action. The 
regulator has to transform as a tensor under gauge transformations of 
$\bar A$, \eq{trafofR}. Here we specify a general class of regulators 
which has this property and is well-suited for practical applications. 
As already argued in section~\ref{sec:absence}, the infrared 
regularisation is provided by $r$, whereas $\tilde r\neq 0$ only 
gives different weights to the longitudinal degrees of freedom, see 
\eq{eq:cut-off2}. In the following we set $\tilde r\equiv 0$. We choose 
\begin{eqnarray}\label{suitable}
R_k[\bar A]= \bar D_T\,r(\bar D_T) 
\end{eqnarray} 
with the yet unspecified function $r$. We introduced 
$D_T$, the Laplace operator for spin 1,
\begin{eqnarray}\label{transversal}
D_{T,\mu\nu}^{ab}(A)  :=  -(D_\rho D_\rho)^{ab}(A) \delta_{\mu\nu}-
2gF_{\mu\nu}^{ab}(A)
\end{eqnarray}  
and $\bar D_T=D_T(\bar A)$. For vanishing background field the
Laplacean $D_T$ reduces to the free Laplacean $D_T(0)=p^2$. In this
case we have $R_k=p^2\, r(p^2)$. 
Written in terms of some general Laplace operator $P^2(\bar A)$, a
typical example for the regulator functions $R_k(P^2)$ and $r(P^2)$ is
\begin{eqnarray}\label{furnish} 
R_k(P^2)={P^2\ov \exp{P^2/k^2}-1}\ ,\quad\quad
r(P^2)={1\ov \exp{P^2/k^2}-1}
\end{eqnarray} 
which meets the general properties as described in \eq{Regulator}.
\Eq{furnish} is an example for a regulator with a mass-like IR
behaviour, $\gamma=1$. More generally the IR/UV conditions for $R_k$ 
in \eq{Regulator} translates into 
\begin{eqnarray}\label{power}
\lim_{k^2/p^2 \to 0} \left(\0{p^2}{k^2}\right)^{2} r =0, 
\qquad \qquad \lim_{P^2\to 0} r \sim  
\left(\frac{k^2}{p^2}\right)^\gamma  
\end{eqnarray}
for the function $r$. \step

\section{Analytic methods}\label{Methods}

In this section we develop analytical methods to study flow equations
for gauge theories in general axial gauges.  The flow equation is 
a one-loop equation which makes it possible to use 
heat kernel techniques for its solution. The main obstacles,
technically speaking, are the constraint imposed by the modified Ward
identity and the necessity to come up with a closed form for the full
propagator. We first derive such an expression for the case of
covariantly constant fields within general axial gauges. In addition a
generic expansion procedure in powers of the fields is discussed.
Finally, we give the basic heat kernels to be employed in the next
section.

\subsection{Propagator for covariantly constant fields}\label{covprop}

We derive an explicit expression for the full
propagator for specific field configurations. This is a prerequisite
for the evaluation of the flow equation \eq{flow}. To that end
 we restrict ourselves to field
configurations with covariantly constant field strength (see
e.g. \cite{McArthur:1997ww}), namely $D_\mu F_{\nu\rho}=0$. This is a common
procedure within the algebraic heat kernel approach.  We also use the
existence of the additional Lorentz vector to demand $n_\mu
A^\mu=n_\mu F_{\mu\nu}=0$. That this can be achieved is proven by the
explicit example of $n_\mu=\delta_{\mu 0}$ and $(A_\mu)=(A_0=0,
A_i(\vec x))$. These constraints lead to 
\begin{mathletters}
\label{zeros}
\begin{eqnarray} \label{covzero}
[D_\mu,F_{\nu\rho}] \tab = \tab 0, \\\di 
n_\mu A_\mu \tab = \tab 0\label{nAzero}\\\di 
n_\mu F_{\mu\nu} \tab = \tab 0. \label{nFzero}
\end{eqnarray} 
\end{mathletters}%
To keep finiteness of the action of such configurations we have to go
to a theory on a finite volume. However, the volume dependence will
drop out in the final expressions and we smoothly can take the limit
of infinite volume. For the configurations satisfying \eq{zeros} we
derive the following properties
\begin{mathletters}\label{property}
\begin{eqnarray}
\label{proper1}
[D^2,D_\mu] \tab = \tab -2g F_{\mu\rho}D_\rho, \\\di 
D_{T,\mu\rho}D_\rho \tab = \tab -D_\mu D^2 \ , 
\label{proper2}\\\di 
[n_\rho D_\rho,D_\mu]\tab =\tab 0
\label{proper3}. 
\end{eqnarray} 
\end{mathletters}%
Defining the projectors $P_n$ and $P_D$ with 
\begin{mathletters}\label{project}
\begin{eqnarray}\label{projectn} 
P_{n,\mu\nu} \tab = \tab {n_\mu n_\nu\ov n^2}, \\\di 
P_{D,\mu\nu} \tab = \tab D_\mu {1\ov D^2} D_\nu \label{projectA}
\end{eqnarray} 
\end{mathletters}%
we establish that 
\begin{equation}
P_D D_T=-P_D D^2 P_D,\qquad P_n D_T=-P_n D^2
\end{equation} 
holds true. 
After these preliminary considerations we consider the gauge-fixed
classical action given in \eq{classical}.We need the propagator on tree
level to obtain the traces on one-loop level. The initial action reads
\begin{equation}\label{GammaLambda}
\Gamma_\Lambda[A] = S_A+S_{\rm gf} \ .
\end{equation}
From \eq{GammaLambda} we derive the full inverse propagator as
\begin{eqnarray}\label{invprop} 
\Gamma_{k,\mu\nu}^{(2)ab}[A,A] = \left(D_{T,\mu\nu}^{ab}+
(D_\mu D_\nu)^{ab}+
\frac{1}{\xi n^2}n_\mu n_\nu\delta^{ab}\right)+O(g^2;D_T,D_\mu D_\nu)\ . 
\end{eqnarray}
The inverse propagator \eq{invprop} is an operator in the adjoint
representation of the gauge group. 
We now turn to the computation of the propagator \eq{G_k} for
covariantly constant fields.  Using \eq{invprop}, \eq{zeros} and
\eq{property}, we find
\begin{eqnarray}\label{propfull} 
\left.G_k[A,A]\right.^{ab}_{\mu\nu} =  
- \left(
 \left(\frac{a_1}{D_T}\right)_{\mu\nu}
+ D_\mu\frac{a_2}{D^4} D_\nu
+ n_\mu\frac{a_3}{D^2 (nD)}D_\nu
+ D_\mu\frac{a_3}{D^2 (nD)}n_\nu
+\frac{ n_\mu  a_4 n_\nu}{n^2 D^2} \right),
\end{eqnarray} 
with the dimensionless coefficient functions 
\begin{mathletters}\label{coeff1} 
\begin{eqnarray} 
a_1 \tab =\tab {1\ov 1+r_T}\ ,\\  
a_2 \tab =\tab {1 - \xi D^2 (1+r_D)\ov(1+r_D)}
         \left( s^2 +r_D[1- D^2\xi (1+r_D)]\right)^{-1}\ ,\\ 
a_3 \tab = \tab 
        - {s^2\ov(1+r_D)}
          \left( s^2 +r_D[1 - D^2 \xi (1+r_D)]\right)^{-1}\ ,\\
a_4 \tab = \tab -{r_D \ov(1+r_D)}
          \left(  s^2 +r_D[1 - D^2 \xi (1+r_D)]\right)^{-1}\ .
\end{eqnarray}
\end{mathletters}%
Notice that $a_1$ is a function of $D_T$ while $a_2$, $a_3$ and $a_4$
are functions of both $D^2$ and $(nD)^2$.  We also introduced the
convenient short-hands
\begin{equation}\\ \label{rTrDs2}
r_T \equiv r_k(D_T),    \qquad  
r_D \equiv r_k(-D^2),   \qquad
s^2 \equiv  {(nD)^2\ov(n^2 D^2)}. 
\end{equation}
The regulator function, as introduced in \eq{suitable}, depends
 on $D_T$. The dependence on $D^2$, as apparent in the
terms $a_2$, $a_3$ and $a_4$, comes into game due to the conditions
\eq{zeros} and \eq{property}. They imply
\begin{equation}
r_k(D_T) D_\mu = D_\mu r_k(-D^2),
\quad\quad 
r_k(D_T) n_\mu = n_\mu r_k(-D^2)\ ,
\end{equation}
which can be shown term by term for a Taylor expansion of $r_k$ about
vanishing argument.  For vanishing field $A=0$ the propagator
\eq{propfull} reduces to the one already discussed in
\cite{Litim:1998qi}. There, it has been shown that the regularised
propagator \eq{propfull} (for $r\neq 0$) is not plagued by the
spurious propagator singularities as encountered within standard
perturbation theory, and in the absence of a regulator term $(r=0)$.
For the axial gauge limit $\xi=0$ the expression \eq{propfull}
simplifies considerably. With \eq{invprop} and \eq{rTrDs2} we get
\begin{eqnarray}\nonumber 
G_{k,\mu\nu}[A]\tab = \tab \!\!  
\left(\frac{1}{D_T(1+r_T)}\right)_{\mu\nu}
-D_\mu\frac{1}{D^4(1+r_D)(s^2+r_D)}D_\nu
+\frac{n_\mu}{n^2}\frac{nD}{D^4(1+r_D)(s^2+r_D)}D_\nu\\ 
\tab &\di  +
D_\mu\frac{nD}{D^4(1+r_D)(s^2+r_D)}\frac{n_\nu}{n^2}
+\frac{r_D}{D^2(1+r_D)(s^2+r_D)}P_{n,\mu\nu}
\ .
\label{fullprop}
\end{eqnarray} 
The propagators \eq{propfull} and \eq{fullprop} are at the basis for
the following computations. Notice that this analysis straightforwardly extends to approximations
for $\Gamma_k[A,\bar A]$ beyond the one-loop level.  Indeed, it applies
for any $\Gamma_k[A,\bar A]$ such that $\Gamma_{k,\mu\nu}^{(2)}[A,A]$
is of the form
\begin{eqnarray}\label{extend} 
\Gamma_{k,\mu\nu}^{(2)}[A,A]= 
\left.     f_{k}^{D_T}              \, D_{T} \right._{\mu\nu} 
+ D_\mu \, f^{DD}_{k}               \, D_\nu
+ n_\mu \, \frac{f^{nD}_{k}}{nD}    \, D_\nu
+ D_\mu \, \frac{f^{nD}_{k}}{nD}    \, n_\nu
+ n_\mu \, f^{nn}_{k}               \, n_\nu\ . 
\end{eqnarray} 
Here, the scale-dependent functions $f_{k}^{D_T}$ and $f^{DD}_{k}$ can
depend on $D_T$, $D^2$ and $nD$.  In turn, the functions $f^{nD}_{k}$
and $f^{nn}_{k}$ can depend only on $D^2$ and $nD$. An explicit
analytical expression for the full propagator, similar to
\eq{propfull}, follows from \eq{extend}. Such approximations take the
full (covariant) momentum dependence of the propagator into account.
The inverse propagator \eq{invprop} corresponds to the particular case
$f^{D_T}=f^{DD}=1$, $f^{nD}=0$, and $f^{nn}=1/\xi $.

\subsection{Expansion in the fields}\label{Expansion} 
Even for analytic calculations one wishes to include more than
covariantly constant gauge fields, and to expand in powers of the
fields, or to make a derivative expansion. Eventually one has to employ
numerical methods where it is inevitable to make some sort of 
approximation. Therefore it is of importance to have a
formulation of the flow equation which allows for simple and
systematic expansions. \step

In this section we are arguing in favour for a different splitting of
the propagator which makes it simple to employ any sort of
approximation one may think of. For this purpose we employ the
regulator $R_k[D^2(\bar A)]$. This is an appropriate choice since it
has no negative eigenvalues.  We split the inverse propagator into
\begin{eqnarray}\label{proprop} 
\Gamma_{k,\mu\nu}^{(2)ab}[A] 
= \Delta_{\mu\nu}^{ab}
 -\left( 2 g F_{\mu\nu}^{ab}- (D_\mu D_\nu)^{ab} \right)
\end{eqnarray} 
with 
\begin{eqnarray}\label{Delta} 
\Delta_{\mu\nu}^{ab}=\left\{-D^2(1+r_D)
\right\}^{ab}\delta_{\mu\nu}+
\frac{1}{\xi n^2}n_\mu n_\nu\delta^{ab}. 
\end{eqnarray} 
The operator $\Delta$ can be explicitly inverted for any field configuration 
(and $A=\bar A$). We have 
\begin{eqnarray}\label{invert2}
\Delta^{-1}= -\frac{1}{D^2(1+r_D)}\id +{1\ov D^2(1+r_D)}{1\ov 1+\xi 
D^2(1+r_D)}P_n. 
\end{eqnarray} 
With \eq{proprop} and \eq{invert2} we can expand the propagator as 
\begin{eqnarray}\label{expand}
G_k[A,A]= \Delta^{-1}\sum_{n=0}^\infty \left[\left(2 g F-D\otimes D\right)
\Delta^{-1}\right]^n. 
\end{eqnarray} 
where $(D\otimes D)^{ab}_{\mu\nu}(x,y)=D_\mu^{ac} D_\nu^{cb}\delta(x-y)$. 
For $\xi=0$ (the axial gauge), $\Delta^{-1}$ can be neatly written as 
\begin{eqnarray}\label{neat} 
\Delta^{-1}(\xi=0)= -{1\ov D^2(1+r_D)} (\id -P_n), 
\end{eqnarray} 
which simplifies the expansion \eq{expand}. 
The most important points in \eq{expand} concern the fact that it is
valid for arbitrary gauge field configurations and each term is
convergent for arbitrary gauge fixing parameter $\xi$. Moreover such
an expansion is not possible in the case of covariant gauges.
Both facts mentioned above are spoiled in this case.

\subsection{Heat kernels}\label{heat-kernel} 
We present closed formulae for the heat-kernel of the
closely related operators $D_T$ and $-D^2=D_T+2gF$. These are needed
in order to evaluate the traces in \eq{bar1loop}.  We define the
heat-kernels as $K_\CO(\tau)=\exp\{\tau \CO\}(x,x)$
\begin{mathletters}\label{HK}
\begin{eqnarray}\label{HKD^2}
K_{D^2}(\tau)   \tab = \tab 
\int\frac{d^4 p}{(2\pi)^4}e^{\tau X_\mu X_\mu},\\\di 
K_{-D_T}(\tau)  \tab =\tab   e^{2 \tau F} K_{D^2}(\tau),\label{HKD_T}
\end{eqnarray}
\end{mathletters}%
where $X_\mu=ip_\mu+D_\mu$ in the corresponding representation.  Here
we used that $2g F$ commutes with $X_\mu$ for covariantly constant
fields.  All kernels are tensors in the Lie algebra ( $K_{-D_T}$ is
also a Lorentz tensor because of the prefactor).  For the calculation
of the momentum integral we just refer the reader to the literature
(e.g.  \cite{McArthur:1997ww}) and quote the result for covariantly constant
field strength
\begin{mathletters}\label{constant}
\begin{eqnarray}
K_{D^2}(\tau)  &= & \frac{1}{16\pi^2 \tau^2}\det\left[\frac{\tau gF}{\sinh 
\tau gF}\right]^{1/2}\ ,
\\  
K_{-D_T}(\tau)    &= &\exp(2 \tau gF)\ K_{D^2}(\tau) \ .
\end{eqnarray} 
\end{mathletters}%
Here, the determinant is performed only with respect to the Lorentz
indices.  For the computation of the one-loop beta function we need to
know $K(\tau)$ in \eq{constant} up to order $F^2$ (equivalently to
order $\tau^0$).  Expanding $K_{D^2}$ in $\tau gF$ we get
\begin{eqnarray}\label{expansion}
K_{D^2}(\tau) = \frac{1}{16\pi^2}
\left( \frac{1}{\tau^2}
      -\frac{1}{12}g^2 (F^2)_{\rho\rho}\right)
      +O[\tau,(gF)^3]. 
\end{eqnarray} 
With \eq{expansion} and the expansion $(\exp 2\tau g F)_{\mu\nu}=
1+2\tau g F_{\mu\nu}+2\tau^2 g^2 (F^2)_{\mu\nu}+O[\tau,(gF)^3]$ we
read off the coefficient of the $K(\tau)$ proportional to $F^2$,
\begin{mathletters}\label{F^2terms}
\begin{eqnarray}
\Tr\, K_{D^2}|_{F^2}&=& -{1\ov 16 \pi^2}\, {4\over 3}N g^2\, S_A[A]  \ ,
\\
\Tr\, K_{-D_T} |_{F^2}&=& {1\ov 16 \pi^2}\, {20\over 3} N g^2\, S_A[A]\ , 
\end{eqnarray} 
\end{mathletters}%
where the trace $\Tr$ denotes a sum over momenta and indices. We have
also used that $S_A[A]={1\ov 2} \int \tr_{\rm f} F^2$ with $ \tr_f\,
t^a t^b =-\s012 \delta^{ab}$. Since the operators $D_T$ and $D^2$ carry
the adjoint representation the trace $\Tr$ includes $\tr_{ad}$ with $2
N \tr_{\rm f} t^a t^b= \tr_{\rm ad} t^a t^b$.

\section{Applications}\label{Applications}
In order to put the methods to work we consider in this section the
full one-loop effective action for $SU(N)$ Yang-Mills theory which
entails the universal one-loop beta function for arbitrary regulator
function.

\subsection{Effective action} 
For the right hand side of the flow we need 
\begin{eqnarray}\label{defofGk}
\Gamma_k[A,\bar A]= {1\ov 2}\int
Z_F(t)\,\tr_f\, F^2(A) +S_{gf}[A]+O[(g A)^5,g^2\partial A], & 
\qquad \tab  \tr_f\, t^a t^b =-\frac{1}{2}\delta^{ab}
\end{eqnarray} 
where $\tr_R$ denotes the trace in the representation $R$, $R=f$
stands for the fundamental representation, $R=ad$ for the adjoint
representation. Only the classical action can contribute to the flow,
as $n$-loop terms in \eq{defofGk} lead to $n+1$-loop terms in the
flow, when inserted on the right hand side of \eq{flowphys}. 
This Ansatz leads to the propagator \eq{fullprop} which together with
our choice for the regulator \eq{suitable} is the input in the flow
equation \eq{flowphys}. 
We also use the following in the evaluation of the different
terms in \eq{flowphys}:
\begin{eqnarray}\label{average}
\tr\, D^2= 4 \tr\, D\otimes D 
\end{eqnarray}
With this we finally arrive at 
\begin{eqnarray}\label{fullflow} 
\partial_t \hat \Gamma_k = 
\frac{1}{2} \Tr \left\{ 
  \frac{\partial_t r(D_T)}{1+r(D_T)}
 -\frac{1}{2}\frac{\partial_t r(-D^2)}{1+r(-D^2)} 
 +\frac{1}{4}\frac{\partial_t r(-D^2)}{s^2+r(-D^2)}
\right\}, 
\end{eqnarray} 
where the trace $\Tr$ contains also the Lorentz trace and the adjoint
trace $\tr_{ad}$ in the Lie algebra. The first term on the right-hand
side in \eq{fullflow} has a non-trivial Lorentz structure, while the
two last terms are proportional to $\delta_{\mu\nu}$. 
We notice that the flow equation \eq{fullflow} is well-defined in both
the IR and the UV region. We apply the heat-kernel results
of section~\ref{heat-kernel} to the calculation of \eq{fullflow}. To
that end we take advantage of the following fact: Given the existence
(convergence, no poles) of the Taylor expansion of a function $f(x)$
about $x=0$ we can use the representation
\begin{eqnarray}\label{represent} 
f(-\CO)=f(-\partial_\tau)\exp\{\tau \CO\}|_{\tau=0}  
\end{eqnarray}
Due to the infrared regulator the terms in the flow equation
\eq{fullflow} have this property, where $\CO=D_T,D^2$. Hence we can
rewrite the arguments $D_T$ and $-D^2$ in \eq{fullflow}
as derivatives w.r.t.\ $\tau$ of the corresponding heat kernels
$K_{-D_T}(\tau)$ and $K_{D^2}(\tau)$. Applying this to the flow
equation \eq{fullflow} we arrive at
\begin{eqnarray}
\nonumber 
\partial_t \hat\Gamma_k&= &
\frac{1}{2}\Biggl[ 
 \frac{\partial_t r(-\partial_\tau)}{1+r(-\partial_\tau)}\,
  \Tr K_{-D_T}(\tau) 
 -{1\ov 2}\frac{\partial_t r(-\partial_\tau)}{1+r(-\partial_\tau)}\,
  \Tr K_{D^2}(\tau)
\\\di \tab \tab 
\quad +{1\ov 4} \int \d p_n
\frac{(p_n^2-\partial_\tau)\partial_t r(p_n^2-\partial_\tau)}{p_n^2+
(p_n^2-\partial_\tau)r(p_n^2-\partial_\tau)}
\frac{\tau^{1/2}}{\sqrt{\pi}}\Tr K_{D^2}(\tau) 
\Biggr]_{\tau=0}
\label{fullflow1}\end{eqnarray}
The two terms in the first line follow from
\eq{fullflow}. The last term is more involved because it depends on
both $D^2$ and $nD$ due to $s^2\equiv (nD)^2/n^2D^2$. We note
that $nD=(n\partial)$ holds for configurations satisfying \eq{covzero}
and only depends on the momentum parallel to $n_\mu$. Furthermore it
is independent of the gauge field.  Now we use the splitting of
$(p_\mu) = (p_n,\vec p)$ where $p_n= P_n p$ and $\vec p=(1-P_n)p$. The
heat kernel related to ${\vec D^2}$ follows from the one for $D^2$
via the relation $K_{\vec
  D^2}(\tau)=\frac{\tau^{1/2}}{\sqrt{\pi}}K_{D^2}(\tau)$ as can be
verified by a simple Gau\ss ian integral in the
$p_n$-direction.\step

With these prerequisites at hand, we turn to the full effective action
at the scale $k$, which is given by
\begin{eqnarray}\label{fulldef}
\hat\Gamma_k = \hat\Gamma_{\Lambda}
           +\int_\Lambda^k d k'\ \frac{\partial \hat\Gamma_{k'}}{\partial k'} \ , 
\end{eqnarray} 
where $\Lambda$ is some large initial UV scale. We
start with the classical action $\Gamma_\Lambda= S_A+S_{\rm gf}$.
Performing the $k$-integral in \eq{fulldef} we finally arrive at
\begin{eqnarray}\nonumber 
\hat\Gamma_k[A] &= & 
\left(1+ \s0{N g^2}{16\pi^2} \left(\s0{22}{3}-7
(1-\gamma)\right)\ln {k}/{\Lambda}\right) S_A[A]
\\ \di 
&&
+S_{\rm gf}[A]
+ \sum_{m=1}^\infty C_m(k^2/\Lambda^2)\ \Delta\Gamma^{(m)}[g F/k^2]
+{\rm const}. 
\label{full1loop}\end{eqnarray}
The combination $S_A+S_{\rm gf}$ on the right-hand side of
\eq{full1loop} is the initial effective action. All further terms stem
from the expansion of the heat kernels
\eq{constant} in powers of $\tau$. The terms $\sim \tau^{-2}$ give
field-independent contributions, while those $\sim \tau^{-1}$ are
proportional to $\tr F$ and vanish. The third term on the right-hand
side of \eq{full1loop} stems from the $\tau^0$ coefficient of the heat
kernel.  This term also depends on the regulator function through the
coefficient $\gamma$ \eq{power}.  All higher order terms $\sim \tau^m,
m>0$ are proportional to the terms $C_m(k^2/\Lambda^2)
\,\Delta\Gamma^{(m)}[gF/k^2]$.  These terms have the following
structure: They consists of a prefactor
\begin{mathletters}\label{HigherM}
\begin{equation}\label{prefactor}
C_m(x)=-{1\ov 4m} {(-)^{m}\ov m!} 
\left(1-x^{m} \right)
\end{equation}
and scheme-dependent functions of the field strength, 
$\Delta\,\Gamma^{(m)}[gF]$, each of which is of the order $2+m$
in the field strength $gF$.   
They are given explicitly as
\begin{equation} \label{m}
\Delta\Gamma^{(m)}[g F]=        
       B^{D_T}_m\                       \Tr \, K^{(m)}_{-D_T}(0)
+\left(B^{D^2}_m\, +\, B^{nD}_m \right) \Tr \, K^{(m)}_{D^2} (0)\ .
\label{Cm}
\end{equation} 
\end{mathletters}%
Here, $K^{(m)}_{D^2} (0)$ and $K^{(m)}_{-D_T}(0)$ denote the expansion
coefficients of the heat kernels. We use the following identity
\begin{eqnarray} 
\ f^{(m)}(0)&=&\left. f(\partial_\tau) \tau^m \right|_{\tau=0} , 
\end{eqnarray} 
and $f^{(m)}(x)=(\partial_x)^m f(x)$. In addition, the terms in \eq{m}
contain the scheme-dependent coefficients
\begin{mathletters}\label{B}
\begin{eqnarray}
B^{D_T}_m &=& \left( {\dot r_1\ov 1+r_1}\right)^{(m)}(0)\ , 
\\ \di
B^{D^2}_m &=& -{1\ov 2}B^{D_T}_m \ ,
\\ \di
B^{nD}_m&=& {(-1)^{m+1}\over 4} \int_0^\infty dx 
\left(\partial_x -{1\ov x}\alpha \partial_\alpha \right)^{m+1} 
\left.{\dot r_1(x)\ov \sqrt{r_1(x)}\sqrt{r_1(x)+\alpha}}\right|_{\alpha=1}\ .
\end{eqnarray}
\end{mathletters}%
The coefficients $B^{D_T}$, $B^{D^2}$ and $B^{nD}$ follow from the
first, second and third term in \eq{fullflow}.  We introduced
dimensionless variables by defining $r_1(x)=r(x k^2)$ and $\dot
r_1(x)\equiv \partial_t r_1(x)=-2 x k^2 r'(xk^2) =-2 x r_1'(x)$, in
order to simplify the expressions and to explicitly extract the
$k$-dependence into \eq{prefactor}. The explicit derivation of
$B^{nD}$ is tedious but straightforward and is given -- together with
some identities useful for the evaluation of the integral and the
derivatives -- in appendix~\ref{AppA}. All coefficients $B^{D_T}$,
$B^{D^2}$ and $B^{nD}$ are finite. The appearance of roots in the coefficient 
$B^{nD}$ is not surprising after the discussion of the absence of 
spurious singularities in section~\ref{sec:absence}. \step 

In particular, we can read off the
coefficients for $m=0$ which add up to the prefactor of the classical
action in \eq{full1loop}:
\begin{eqnarray}\label{m=0}
B_0^{D_T}=2 \gamma, \qquad B_0^{D^2}=- \gamma, \qquad B_0^{nD}= 
-{1\over 2}(1-\gamma),  
\end{eqnarray}
where we have used \eq{term3} in the appendix. 
Together with the heat kernel terms proportional to $\tau^0$
given in \eq{F^2terms} this leads to \eq{full1loop}. \step

This application can be extended to include non-perturbative
truncations. The flow of the coefficients \eq{Cm} becomes non-trivial,
and regulator-dependent due to the regulator-dependence of the
coefficients \eq{B}. Then, optimisation conditions for the flow can be
employed to improve the truncation at hand \cite{Litim:2000ci}.\step

Finally, we discuss the result \eq{full1loop} in the light of the
derivative expansion.  Typically, the operators generated along the
flow have the structure $F\, f_k[(D^2+k^2)/\Lambda^2]\,F$, and similar
to higher order in the field strength. For dimensional reasons, the
coefficient function $f_k(x)$ of the operator quadratic in $F$
develops a logarithm $\sim \ln x$ in the infrared region. An
additional expansion of this term in powers of momenta leads to the
spurious logarithmic infrared singularity as seen in \eq{full1loop}.
To higher order in the field strength, the coefficient function behave
as powers of $1/(D^2+k^2)$, which also, at vanishing momenta, develop
a spurious singularity in the IR, and for the very same reason. All
these problems are absent for any finite external gluon momenta, and
are an artifact of the derivative expansion.  A second comment
concerns the close similarity of \eq{full1loop} with one-loop
expressions found within the heat-kernel regularisation. In the latter
cases, results are given as functions of the proper-time parameter
$\tau$ and a remaining integration over $d\ln \tau$.  Expanding the
integrand in powers of the field strength and performing the final
integration leads to a structure as in \eq{full1loop}, after
identifying $\tau\sim k^{-2}$.  In particular, these results have the
same IR structure as found in the present analysis.

\subsection{Running coupling}\label{1loop} 
We now turn to the computation of the beta function at one loop. 
We prove that the result is independent of the choice of the regulator 
and agrees with the standard one. However, it turns out 
that the actual computation depends strongly on the 
precise small-momentum behaviour of the regulator, which makes a 
detailed discussion necessary. \step 

Naively we would read-off the $\beta$-function from the $t$-running of
the term proportional to the classical action $S_A$ in \eq{full1loop}.
Using \eq{nonrenorm} leads to $\partial_t \ln Z_g=-\s012\partial_t \ln
Z_F$. We get from \eq{full1loop}
\begin{eqnarray}\label{rewrite}
Z_F=  \left(\s0{22}{3}-7
(1-\gamma)\right){N g^2\over 16\pi^2} t 
\tab \quad \rightarrow\quad \tab \partial_t \ln Z_g = 
-\left(\s0{11}{3}-\s0{7}{2}
(1-\gamma)\right)
{N g^2\over 16\pi^2}+O(g^4).
\end{eqnarray}
We would like to identify $\beta=\partial_t \ln Z_g$. This relation,
however, is based on the assumption that at one loop one can trade the
IR scaling encoded in the $t$-dependence of this term directly to a
renormalisation group scaling. This assumption is based on the
observation that the coefficient of $S_A[A]$ is dimensionless and at
one loop there is no implicit scale dependence. It is the latter
assumption which in general is not valid. A more detailed analysis of
this fact is given in \cite{Pawlowski:2001df}.  Here, we observe that
the background field dependence of the cut-off term inflicts 
contributions to $\partial_t Z_F S_{\rm cl}$. These terms would be
regulator-dependent constants for a standard regulator without $\bar A$. 
As mentioned below
\eq{deltaS}, one should see the background field as an index for a
family of different regulators. We write the effective action as  
\begin{eqnarray}\label{splitting}
\Gamma_k[A,\bar A]=\Gamma_{k,1}[A]+\Gamma_{k,2}[\bar A]+
\Gamma_{k,3}[A,\bar A]\ .
\end{eqnarray} 
The second term only depends on $\bar A$ and is solely related to the
$\bar A$-dependence of the regulator.  The last term accounts for
gauge invariance of $\Gamma_k$ under the combined transformation
$\del_\omega+\bar\del_\omega$. This term vanishes in the present
approximation, because of the observation that our Ansatz is invariant
-- up to the gauge fixing term -- under both $\del_\omega$ and
$\bar\del_\omega$ separately. The physical running of the coupling is
contained in the flow of $\Gamma_{k,1}[A]$. This leads to
\begin{eqnarray}\label{projectuv}
\beta= -\s012 \partial_t Z_F +\s012 \partial_t Z_{F,2}, 
\end{eqnarray}
where $Z_{F,2}$ is the scale dependence of $\Gamma_{k,2}\propto
Z_{F,2} S_A[A]$.  We rush to add that this procedure is only necessary
because we are interested in extracting the universal one-loop
$\beta$-function from the flow equation. For integrating the flow
itself this is not necessary since for $k=0$ the background field
dependence disappears anyway.  For calculating $\partial_t \ln
Z_{F,2}$ we use \eq{fullprop} and \eq{average} and get
\begin{eqnarray}\nonumber
\partial_t 
\frac{\delta }{\delta \bar A_\mu^a}\Gamma_k[A,\bar A=A] \tab = &\di\!\! 
\frac{1}{2}
\Tr \partial_t \Biggl\{ \frac{ R'_k[D_T]}{D_T+
R_k[D_T]}\frac{\delta D_T}{\delta \bar A_\mu^a}+
{1\ov 2}\frac{R'_k(-D^2)}{-D^2+ R_k[-D^2]}
\frac{\delta D^2}{\delta \bar A_\mu^a} 
\\\di 
\tab &\di 
-{1\ov 4}\frac{R_k'[-D^2]}{(-nD)^2+R_k[-D^2]}
\frac{\delta D^2}{\delta \bar A_\mu^a}\Biggr\}, 
\label{bar1loop} \end{eqnarray} 
where we have introduced the abbreviation 
\begin{eqnarray}\label{abbrev} 
R_k'(x)=\partial_x R_k(x).
\end{eqnarray}
For the derivation of \eq{bar1loop} one uses the cyclycity of the
trace and the relations \eq{property}. We notice that \eq{bar1loop} is
well-defined in both the IR and the UV region.  The explicit
calculation is done in appendix~\ref{AppB}. Collecting the results
\eq{barterm1},\eq{barterm2},\eq{barterm3} we get
\begin{eqnarray}\label{contbar} 
\partial_t \delta_{\bar A} \Gamma_k[A,\bar A=A]|_{F^2}=  
-{ N g^2 \ov 16 \pi^2 }\, 7(1-\gamma)\,\delta_{A} S_A[A]\tab 
\rightarrow\tab \partial_t Z_{F,2}=-
{N g^2\ov 16 \pi^2 }\, 7 (1-\gamma)
\end{eqnarray} 
We insert the results \eq{rewrite} for $\partial_t Z_F$ and 
\eq{contbar} for $\partial_t Z_{F,2}$ in \eq{projectuv} 
and conclude  
\begin{eqnarray}\label{betaUV}
\beta  =  - {11\over 3}\frac{N g^2}{16 \pi^2}+O(g^4). 
\end{eqnarray} 
which is the well-known one-loop result. For regulators with a
mass-like infrared limit, $\gamma=1$, there is no implicit scale
dependence at one loop.  It is also worth emphasising an important
difference to Lorentz-type gauges within the background field
approach. In the present case only the {\it physical} degrees of
freedom scale implicitly with $t=\ln k$ for $\gamma\neq 0$. This can
be deduced from the prefactor $7(1-\gamma)$ in \eq{contbar}. Within
the Lorentz-type background gauge, this coefficient is
$\frac{22}{3}(1-\gamma)$ \cite{Pawlowski:2001df}.  The difference has
to do with the fact that in the axial gauge one has no auxiliary
fields but only the physical degrees of freedom. In a general gauge,
this picture only holds true after integrating-out the ghosts. This
integration leads to non-local terms. They are mirrored here in the
non-local third term on the right hand side of the flow \eq{fullflow1}
and in the third term on the right hand side of \eq{bar1loop} [see
also \eq{barterm3}].

\section{Conclusions} \label{Discussion}

We have shown how the exact renormalisation group can be used for
gauge theories in general axial gauges. We have addressed various
conceptual points, in particular the absence of spurious singularities
and gauge invariance, which are at the basis for a reliable
application of this approach. We have shown that spurious
singularities are absent provided that the regulator $R_k$ decays
stronger than $(p^2)^{-4}$ for large momenta. In turn, regulators with
milder decay are highly questionable. At least they are subject to a
renormalisation of the flow itself, which implicitly brings back the
problem of spurious singularities. This concerns in particular the
mass regulator $R_k=k^2$, see also \cite{Litim:1998nf}. \step

Our main goal was to develop methods
which allow controlled and systematic analytical considerations.  The
formalism has the advantage that ghost fields are not required. Also,
no additional regularisation -- in spite of the axial gauge fixing --
is needed.  This is a positive side effect of the Wilsonian regulator
term. In addition, we worked in a background field formulation, which is
helpful in order to construct a gauge invariant effective action.
Also, it allows to expand the flow equation around relevant field
configurations. Instead of relying on the standard background field
gauge, we have introduced the background field only in the regulator
term. The axial gauge fixing is independent on the background field.
This way, it is guaranteed that the background field dependence
vanishes in the IR limit.  It is important to discuss how this differs
from the usual background field approach to Wilsonian flows. In both
cases, applications of the flow require an approximation, where
derivatives w.r.t.\ the background field are neglected,
cf.~\eq{approx}. In the present approach, this approximation improves
in the infrared, finally becoming exact for $k=0$ as the background
field dependence disappears. For the background field gauge this does
not happen, because the full effective action still depends
non-trivially on the background field.  \step

As an application, the full one-loop effective action and the
universal beta-function have been computed. This enabled us to address
some of the more subtle issues of the formalism like the implicit
scale dependence introduced by the cutoff, which has properly to be
taken into account for the computation of universal quantities,
and the scheme independence of the beta-function. The
equation which controls the additional background field dependence
introduced by the cutoff contains the related information. \step

These results are an important step towards more sophisticated
applications, both numerically and analytically. A natural extension
concerns dynamical fermions. The present formalism is also
well-adapted for QCD at finite temperature $T$, where the heat-bath
singles-out a particular Lorentz vector. Here, an interesting
application concerns the thermal pressure of QCD.\step

\section*{Acknowledgements}
We thank P.~Watts for helpful discussions. JMP thanks CERN for
hospitality and financial support.  DFL has been supported by the
European Community through the Marie-Curie fellowship
HPMF-CT-1999-00404.\step

\setcounter{section}{0}
\renewcommand{\thesection}{\Alph{section}}
\renewcommand{\thesubsection}{\arabic{subsection}}
\renewcommand{\theequation}{\Alph{section}.\arabic{equation}}

\setcounter{equation}{0}
\section{Evaluation of the one loop effective action}\label{AppA}
The calculation of the last term in \eq{full1loop} is a bit more
involved. Note that the following argument is valid for $m\geq -1$,
$m>-1$ is of importance for the evaluation of \eq{full1loop}, $m=-1$
will be used in Appendix~\ref{AppB}.  We first convert the factor
$\tau^{m+1/2}$ appearing in the expansion of the heat kernel using
$\tau^{1/2+m}=(-1)^{m+1} {\tau\ov\sqrt{\pi}}\int d z
\partial_{z^2}^{m+1} e^{-\tau z^2}$.  We further conclude that
\begin{eqnarray}\nonumber 
B_m^{n D}\tab= \tab 
{1\over 4\pi}\int d p_n\,d z {(p_n^2-\partial_\tau)
\partial_t r(p_n^2 -\partial_\tau)\ov p_n^2+
(p_n^2-\partial_\tau) r(p_n^2-\partial_\tau)} \tau^{m+1} e^{-\tau z^2}
|_{\tau=0}\\\di \nonumber 
\tab =\tab {(-1)^{m+1}\over 4\pi} \int d p_n\,d z 
\left. \partial^{m+1}_{z^2} 
{\partial_t r(p_n^2 -\partial_\tau)\ov p_n^2+
(p_n^2-\partial_\tau)
r(p_n^2-\partial_\tau)}(p_n^2-\partial_\tau) e^{-\tau z^2}
\right|_{\tau=0}\\\di 
\tab =\tab {(-1)^{m+1}\over 4\pi} \int d p_n d z 
\partial^{m+1}_{z^2} {\partial_t r(z^2+p_n^2)\ov {p_n^2\ov z^2+p_n^2}+
r(z^2+p_n^2)},  
\label{step1}\end{eqnarray} 
The expression in \eq{step1} 
can be conveniently rewritten as 
\begin{eqnarray}\nonumber
B_m^{n D}\tab= \tab   {(-1)^{m+1}\over 8\pi}
\int_0^\infty dx \int_0^{2\pi} d\phi 
\left. \left(\partial_x -{1\ov x}\alpha\partial_\alpha \right)^{m+1}
{\partial_t r(x)\ov \alpha\sin^2\phi+r(x)}\right|_{\alpha=1}\\\di 
\tab =\tab {(-1)^{m+1}\over 4}\int_0^\infty dx 
\left. \left(\partial_x -{1\ov x}\alpha\partial_\alpha \right)^{m+1} 
{\partial_t r(x)\ov 
\sqrt{r(x)}\sqrt{r(x)+\alpha}}\right|_{\alpha=1}. 
\label{step2}\end{eqnarray} 
where $x=z^2+p_n^2$ and $\sin^2\phi=p_n^2/(z^2+p_n^2)$. 
It is simple to see that 
$-(1/x)\alpha\partial_\alpha$ is a representation of $\partial_{z^2}$ 
on $\sin^2\phi=p_n^2/(z^2+p_n^2)$ and $\partial_x$ a representation of 
$\partial_{z^2}$ on functions of $x$ only. 
The expression in \eq{step2} is finite for all $m\geq 0$. Evidently it falls 
of for $x\to \infty$. For the behaviour at $x=0$ the 
following identity is helpful:
\begin{eqnarray}\label{binomial}
\left(\partial_x -{1\ov x}\alpha\partial_\alpha \right)^{m+1}=\sum_{i=0}^{m+1}
(-1)^{m+1-i}
\renewcommand{\arraystretch}{1}
\left(
\begin{array}{c}\di  
m+1 \\\di i 
\end{array}\right)
\renewcommand{\arraystretch}{1.9}
\partial_x^i \left({\alpha\ov x}\right)^{m+1-i} 
\partial_\alpha^{m+1-i}, 
\end{eqnarray} 
\Eq{binomial} guarantees that the integrand in 
\eq{step2} only contains terms of the form 
\begin{eqnarray}\label{form} 
\partial^i_x\left( {\dot r\ov \sqrt{r}\sqrt{1+r}} (x+x r)^{i-m-1}\right)
\end{eqnarray} 
with $i=0,...,m+1$. For $x\to 0$ one has to use that $\partial_t r \to 2 n 
r$ and $r\to {k^{2n}\ov x^n}$. The terms of integrand in \eq{step2} as 
displayed in \eq{form} are finite for $x=0$. 

We are particularly interested in $B_0^{nD}$ relevant for 
the coefficient of $S_A$ in the one loop effective action \eq{full1loop}. 
With \eq{step2} it follows   
\begin{eqnarray} \nonumber 
B_0^{nD}\tab =\tab  -{1\over 4}\int_0^\infty dx 
\left. \left(\partial_x -{1\ov x}\alpha\partial_\alpha \right) 
{\partial_t r(x)\ov 
\sqrt{r(x)}\sqrt{r(x)+\alpha}}\right|_{\alpha=1}\\\di  
\tab =\tab -{1\over 4}\left({\partial_t r(x)\ov 
\sqrt{r(x)}\sqrt{1+r(x)}}-
2 {\sqrt{r(x)}\ov \sqrt{1+r(x)}}\right)_{x=0}^{x=\infty}= -{1\over 2}
(1-\gamma),  
\label{term3}\end{eqnarray} 
where we have used $\partial_t r(z)=-2 z\partial_z r(z)$ and the  
limits for $\partial_t r(z\rightarrow 0)= 2 \gamma z^{-\gamma}, 
r(z\rightarrow 0)= z^{-\gamma}, r(z\rightarrow\infty)=0$.  
\renewcommand{\theequation}{B.\arabic{equation}}
\setcounter{equation}{0}
\section{$\bar A$-Derivatives}\label{AppB}
For the calculation of \eq{bar1loop} the following identity is useful: 
\begin{eqnarray}\label{barderiv}
\Tr\left(\frac{\delta}{\delta A_\mu^a} {\cal O}\right)  
e^{\tau {\cal O}}=\frac{1}{\tau}\Tr
\frac{\delta}{\delta A_\mu^a} e^{\tau {\cal O}},  
\end{eqnarray}  
where we need \eq{barderiv} for ${\cal O}=D^2$ and ${\cal O}=-D_T$. 
Now we proceed in calculating the first term in \eq{bar1loop} by using a 
similar line of arguments as in the calculation of \eq{full1loop} and 
in Appendix~\ref{AppA}. We make use of the representation of 
$\tau^{-1}=\int_{0}^\infty dz \exp{-\tau z}$ and arrive at  
\begin{eqnarray}\nonumber 
\frac{1}{2}
\Tr\,\partial_t\left(\frac{R_k'[D_T]}{D_T+
R_k[D_T]}\frac{\delta D_T}{\delta A_\mu^a} \right)  \tab = \tab
\frac{1}{2}
\Tr\,\partial_t\left(\frac{R_k'(-\partial_\tau)}{-\partial_\tau+
R_k[-\partial_\tau]}\frac{1}{\tau}\frac{\delta}{\delta A_\mu^a}
K_{-D_T}(\tau)\right)_{\tau=0}
\\\di \nonumber 
\tab = &\di 
{1\ov 2}\int_0^\infty {\d x\over x} \partial_t
\left(\frac{R_k'[x]}{1+r[x]}\right)
 {N g^2 \ov 16\pi^2} {20\ov 3} 
\frac{\delta}{\delta A_\mu^a}\left( S_A[A]
+O[g]\right)\\\di 
\tab =\tab 
-{N g^2\ov 16\pi^2}\,{20\ov 3}\,(1-\gamma)\frac{\delta}{\delta  A_\mu^a}
 \left(S_A[A]+O[g]\right). 
\label{barterm1}\end{eqnarray}  
Note that $\partial_t$ acts as $-2 x\partial_x$ on functions which
solely depend on $x/k^2$. The term $R'/(1+r)$ is such a function.  The
second term can be calculated in the same way leading to
\begin{eqnarray}\nonumber 
\frac{1}{4}
\Tr \ \partial_t \left\{\frac{-R_k'[D^2]}{-D^2+
R_k[-D^2]}\frac{\delta}{\delta A_\mu^a} D^2\right\} 
\tab =\tab  \frac{1}{4}
\int_0^\infty{\d x\over x}\ \partial_t 
\left(\frac{R_k'[x]}{1+r[x]}\right)
 {N g^2 \ov 16\pi^2} {4\over 3}
\frac{\delta}{\delta  A_\mu^a}\left( S_A[A] +O[g]\right)
\\\di 
\tab =\tab 
-{N g^2\ov 16\pi^2}\,{2\over 3}\,  (1-\gamma)
\frac{\delta}{\delta  A_\mu^a}\left(
S_A[A]+O[g]\right). 
\label{barterm2}\end{eqnarray} 
The  calculation of  the  last term  in  \eq{bar1loop} is  a bit  more
involved, but boils down to the same structure as for the other terms.
Along the lines  of Appendix~\ref{AppA} it follows that  this term can
be written as
\begin{eqnarray}\nonumber 
{1\ov 8}\Tr\,\partial_t\,\Biggr\{
\frac{ -R_k'[-D^2]}{(-nD)^2+R_k[-D^2]}
\frac{\delta D^2}{\delta A_\mu^a}\Biggr\}
\tab  =\tab 
\frac{1}{8}\Tr\ \partial_t\left\{ 
\int \d p_n  \frac{R_k'[p_n^2-\partial_\tau]}{p_n^2+
R_k[p_n^2-\partial_\tau]}
\frac{\tau^{-1/2}}{\sqrt{\pi}}\frac{\delta}{\delta A_\mu^a} 
K_{D^2}(\tau)\right\}_{\tau=0}\\\di\nonumber  
\tab= \tab  
-\frac{1}{8}\int_0^\infty {\d x\ov x} 
 \partial_t \frac{R_k'}{\sqrt{r}\sqrt{1+r}}
 {N g^2\ov 16\pi^2} {4\over 3} 
\frac{\delta }{\delta A_\mu^a}\left( S_A[A]+O[g]
\right),
\\\di 
\tab  =\tab  {N g^2 \ov 16\pi^2}\,{1\over 3}\, 
(1-\gamma)\frac{\delta }{\delta A_\mu^a}
\left(S_A[A]+O[g]\right)
\label{barterm3}\end{eqnarray} 
Note that when rewriting the left hand side of \eq{barterm3} as a total
derivative w.r.t.\ $A$ this also includes a term which stems from
${\delta\ov \delta A} (nD)^2$. This, however, vanishes because it is
odd in $p_n$.
\end{document}